\documentclass[aps,prb,floatfix,amsmath,amssymb,preprint,eqsecnum,superscriptaddress]{revtex4}
\usepackage{graphicx}
\usepackage{dcolumn}
\usepackage{bm}
\usepackage{epstopdf}
\usepackage{enumerate}
\usepackage{setspace}   
\usepackage{amsmath}
\usepackage{amssymb}

\preprint{NSF-KITP-09-129}

\begin{document}
\title{Fluctuating spin density waves in metals}

\author{Subir Sachdev}
\affiliation{Department of Physics, Harvard University, Cambridge MA
02138}
\affiliation{Kavli Institute for Theoretical Physics, University of California,
Santa Barbara CA 93106}

\author{Max A. Metlitski}
\affiliation{Department of Physics, Harvard University, Cambridge MA
02138}

\author{Yang Qi}
\affiliation{Department of Physics, Harvard University, Cambridge MA
02138}

\author{Cenke Xu}
\affiliation{Department of Physics, Harvard University, Cambridge MA
02138}
\affiliation{Kavli Institute for Theoretical Physics, University of California,
Santa Barbara CA 93106}

\date{\today\\
\vspace{1.6in}}
\begin{abstract}
Recent work has used a U(1) gauge theory to describe the physics of Fermi pockets in the
presence of fluctuating spin density wave order. We generalize this theory to an arbitrary
band structure and ordering wavevector. The transition to the large Fermi surface state, without pockets
induced by local spin density wave order, is described
by embedding the U(1) gauge theory
in a SU(2) gauge theory. The phase diagram of the SU(2) gauge theory shows that the onset of 
spin density wave order in the Fermi liquid occurs either directly, in the framework discussed by Hertz,
or via intermediate non-Fermi liquid phases with Fermi surfaces of fractionalized excitations. We discuss 
application of our results to the phase diagram
of the cuprates.
\end{abstract}

\maketitle

\section{Introduction}
\label{sec:intro}

Recent experimental advances \cite{doiron,cooper,nigel,cyril,suchitra,louis,suchitra2} have focused much theoretical attention on the evolution of the
Fermi surfaces of the cuprate superconductors as a function of carrier concentration. In materials
with hole density $x$, the overdoped regime has a ``large'' hole-like Fermi surface enclosing area
proportional to $1+x$, while the underdoped regime has displayed evidence for ``small'' Fermi pockets
with an area of order $x$. We refer the reader to other recent discussions \cite{moon,qcnp} for an overview of the
experimental and theoretical situation suited for the ideas presented below. We show here in Fig.~\ref{figglobal} the
global phase diagram from Ref.~\onlinecite{qcnp} as a function of $x$, temperature $T$, and applied magnetic field $H$.
\begin{figure}
\includegraphics*[width=5.5in]{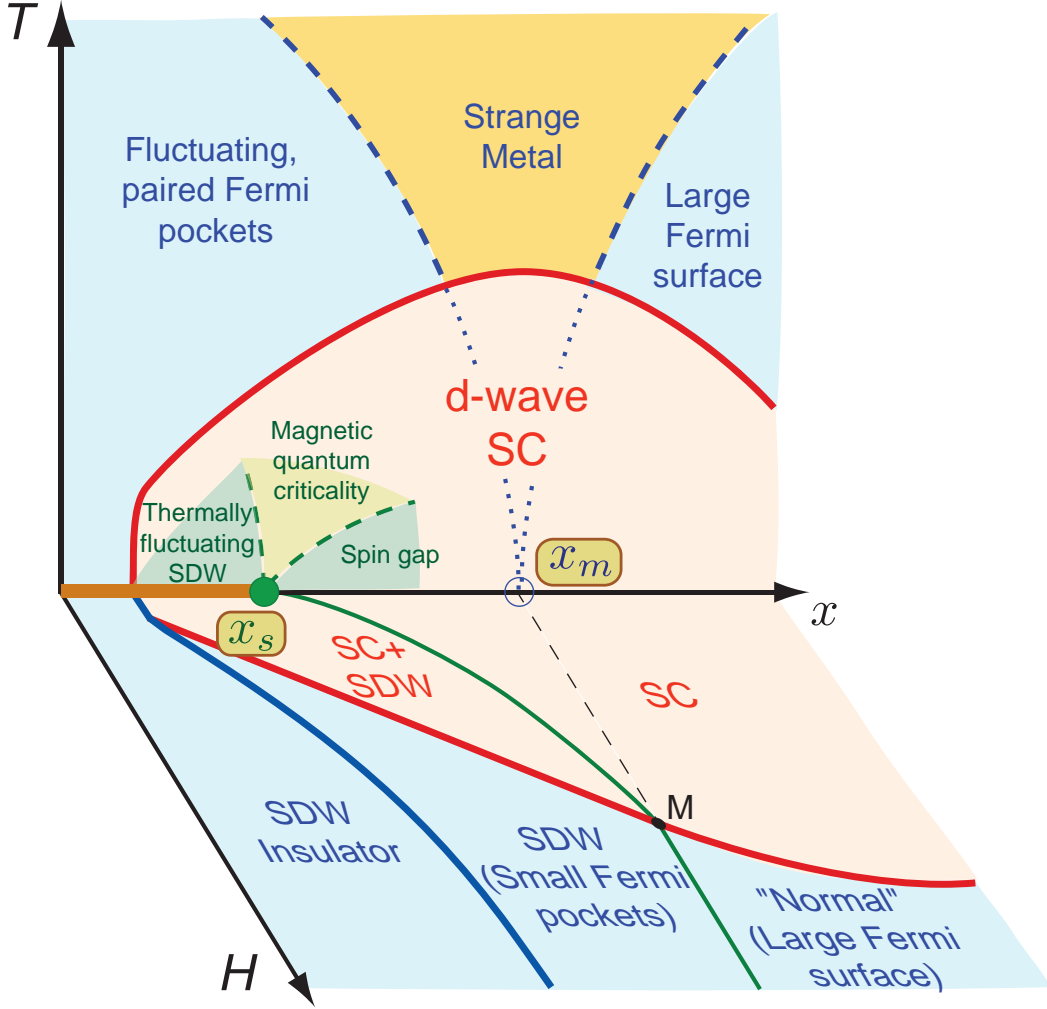}
\caption{From Refs.~\onlinecite{qcnp,moon}: Proposed phase diagram as a function of dopant density $x$, temperature, $T$,
and magnetic field $H$. The onset of long-range 
spin density wave (SDW) order at $T=0$ and high fields in the metallic state is at $x=x_m$, while SDW order appears at $x=x_s$
in the superconducting (SC) state at $H=0$. A key feature of this phase diagram, and of our theory \cite{moon},
is that $x_s < x_m$. This implies the phase transition line connecting $x_s$ and $x_m$, predicted in Ref.~\onlinecite{demler}, where there is a field-induced onset of SDW order in the SC state, which has been experimentally detected \cite{khaykovich,mesot,mesot2,mesot3}.}
\label{figglobal}
\end{figure}

We will be interested in the physics of the non-superconducting metallic phases in Fig.~\ref{figglobal}, when the superconductivity
is suppressed by increasing either $T$ or $H$.
As is implied by Fig.~\ref{figglobal},
we will assume \cite{fuchun,millisnorman,harrison} that the ``small'' Fermi surfaces are a consequence of local spin density
wave (SDW) order: this is supported by a number of recent experiments \cite{mesot,mesot2,mesot3,nlsco1,nlsco2,nlsco3}. 
It is then natural to develop a theory of the electronic spectrum in presence of (thermal or quantum) fluctuating SDW order.
We would like the electronic spectrum to be sensitive to the presence of SDW order at short scales, even though long-range SDW
order can be absent.

A U(1) gauge-theoretic approach to describing such a fluctuating SDW state has been presented by some
of us and our collaborators in a series of papers \cite{rkk1,rkk2,rkk3,gs,moon}. While 
this theory has
a number of attractive features \cite{qcnp}, it also has some weaknesses:
\begin{enumerate}
\item The theory addresses the physics of the small Fermi pocket states only, and is not connected
to the large Fermi surface state of the overdoped regime.
\item The pockets are described in a piecemeal fashion, with separate fermion degrees of freedom introduced
at the band minimum of each pocket. A unified formalism which treats all pockets together, for an arbitrary
underlying band structure would clearly be preferred.
\item The theory has so far focused on commensurate SDW order with ordering wavevector
${\bm K} = (\pi, \pi)$. It should be generalized to arbitrary commensurate ${\bm K}$.
\end{enumerate}
The purpose of the present paper is to present an improved formalism which addresses the above issues.
We will begin in Section~\ref{sec:u1} by a reformulation of the existing U(1) gauge theory which addresses points 2
and 3 above. Section~\ref{sec:su2} will address point 1 by showing that the transition to the large Fermi surface
state is achieved by generalizing to a SU(2) gauge theory. We note that our SU(2) gauge theory is quite distinct from
that appearing in the discussion of spin liquid Mott insulators with fermionic spinons \cite{leewen} in which the SU(2) gauge transformation
mixes and particle and hole operators.
Our theory applies
to bosonic spinons in metals, and the SU(2) gauge transformations apply on states with fixed particle number.

Although our primary motivation has been provided by cuprate physics, our approach is very general, and should
also be applicable to a wide variety of spin density wave transitions in other correlated electron materials \cite{si}.

\section{U(1) gauge theory}
\label{sec:u1}

We begin with the popular spin-fermion model \cite{spinfermion}, for a system where the spins have collinear ordering at an arbitrary
commensurate wavevector ${\bm K}$. The imaginary time ($\tau$) fermion Lagrangian is $(\alpha, \beta = \uparrow, \downarrow )$
\begin{equation}
\mathcal{L}_F = \sum_{i} c^{\dagger}_{i \alpha} \Bigl[(\partial_\tau - \mu) \delta_{\alpha\beta} - \varphi^a_i f_i \sigma^a_{\alpha\beta} \Bigr]
c_{i \beta}
- \sum_{i<j} t_{ij} \left( c^{\dagger}_{i \alpha} c_{j \alpha} + c^{\dagger}_{j \alpha} c_{i \alpha} \right) \label{hsdw}
\end{equation}
Here $t_{ij}$ are arbitrary hopping matrix elements describing the ``large'' Fermi surface, $\mu$ is the chemical potential, 
$\varphi^a_i$ is a fluctuating
unit vector field ($a=x,y,z$) representing the local orientation of the collinear spin order, $\sigma^a$ are the Pauli
matrices, and $f_i$ is a fixed form-factor determined by the particular local nature
of the SDW order; thus for N\'eel order with ${\bm K} = (\pi, \pi)$ we have $f_i \sim (-1)^{x_i + y_i}$, while for arbirary
commensurate ${\bm K}$ we have an expression of the form
\begin{equation}
f_i = \sum_m f_m e^{i m \, {\bm K} \cdot {\bm r}_i} + \mbox{c.c.}
\end{equation}
where $m$ are integer, and the $f_m$ are the co-efficients determining the form-factor.
The fluctuations of the $\varphi^a$ are controlled by the continuum O(3) non-linear sigma model with Lagrangian
\begin{equation}
\mathcal{L}_n = \frac{1}{2g} \left[ (\partial_\tau \varphi^a )^2 + v^2 ({\bm \nabla} \varphi^a)^2 \right] \label{ln}
\end{equation}
with the local constraint $(\varphi^a)^2 = 1$; here $g$ is a coupling which tunes the strength of the quantum spin
fluctuations and $v$ is a spin wave velocity. The spin-fermion model \cite{spinfermion} is defined by the Lagrangian
$\mathcal{L}_F + \mathcal{L}_n$ for the electrons $c_{i\alpha}$ and the SDW order parameter field $\varphi^a$.

We have assumed above that $\varphi^a$ is a real three-component vector. Strictly speaking, for ${\bm K} \neq (\pi,\pi)$, the order
parameter is a complex vector, with the overall phase representing a `sliding' degree of freedom associated with the charge
density wave at $2 {\bm K}$. Indeed, there will be 2 complex vectors representing the spin density waves along two orthogonal
directions on the square lattice. For simplicity, we have ignored these complications here. Accounting for them would require
two additional complex fields, as in {\em e.g.\/} Ref.~\onlinecite{morinari}, and we leave this generalization to future work.

A key feature of our analysis above, and of all the analyses below, is that we assume that it is only the SDW order parameter $\varphi^a$
which varies slowly on the lattice scale. We do not make the same
assumption for the fermion field $c_{i\alpha}$, which is allowed to have a general dispersion, with arbitrary Fermi surfaces.
Thus our expansions in spatial gradients will only be carried out for $\varphi^a$ and related bosonic fields. Keeping the full spatial dependence
of the fermion fields is also required to keep proper track of the constraints imposed by the Luttinger theorem.

We will now transform the spin-fermion model into new degrees of freedom which incorporate the change in the fermion band structure
due to local SDW order in a more fundamental way. The key to doing this is to transform the electron spin polarization to 
a rotating reference frame set by the local orientation of the SDW order. In the context of the cuprates, the use of such a frame
of reference goes back to the work of Shraiman and Siggia \cite{shraiman} on the $t$-$J$ model, and by Schulz \cite{schulz}
on the Hubbard model. Previous work by us and others\cite{rkk1,rkk2,rkk3,gs,moon,wenholon,leeholon,shankar,ioffew}
was motivated using the Schwinger boson formalism, which also effectively performs the transformation to the rotating reference
frame. A few years ago, Schrieffer \cite{schrieffer} 
also focused attention on the advantages of studying the spin dynamics in the rotating reference
frame defined by the SDW order. Here we shall apply this idea to the spin-fermion model, which is generally regarded as
a weak-coupling theory. We shall show that it allows for a very efficient and complete derivation of the Lagrangian of a low energy
effective gauge theory, which has the same structure as that obtained earlier \cite{rkk1,rkk2,rkk3,gs,moon} by more
cumbersome methods starting from the strong-coupling $t$-$J$ model.

To this end, we introduce a new set of fermions, $\psi_{ip}$ with $p =\pm 1$, with their spin components $p$
polarized along the direction of the local SDW order. These are related to the physical fermions $c_{i \alpha}$ by
spacetime dependent SU(2) matrix
$R_{\alpha p}^i$ ($R^\dagger R =  R R^\dagger = 1$) so that \cite{schulz}
\begin{eqnarray}
c_{i \alpha} &=& R_{\alpha p}^i \psi_{i p} . \label{cab}
\end{eqnarray}
We choose $R_{\alpha p}$ so that spin-fermion coupling is only along $\sigma^z$, and so
\begin{equation}
\varphi^a_i R^{i \dagger}_{p \alpha} \sigma^a_{\alpha\beta} R_{\beta p'}^i =  \sigma^z_{p p'} = p \delta_{p p'}.
\end{equation}
This relationship is equivalent to 
\begin{equation}
\varphi^a_i = \frac{1}{2} \mbox{Tr} \left( \sigma^a R^i \sigma^z R^{i\dagger} \right), \label{rz}
\end{equation}
which shows that the SDW order parameter $\varphi_i$ can be fully expressed in terms of 
the SU(2) matrix $R$. Therefore, we will now treat $R$ as our independent degree of freedom
which determines $\varphi$ via Eq.~(\ref{rz}).
Now, we parameterize
\begin{equation}
R^i = \left( \begin{array}{cc} z_{i\uparrow} & -z_{i\downarrow}^\ast \\
z_{i\downarrow} & z_{i\uparrow}^\ast \end{array} \right)
\end{equation}
with $\sum_\alpha |z_{i \alpha} |^2 = 1$, and 
we can verify that Eq.~(\ref{rz}) yields the familiar relation
\begin{equation}
\varphi^a_i =  z_{i\alpha}^\ast \sigma^a_{\alpha\beta} z_{i \beta} \label{nz}
\end{equation}
between the fields of the O(3) non-linear sigma model and the CP$^1$ model.

We have now reformulated our theory of the spin-fermion by replacing the electrons $c_\alpha$
and SDW order parameter $\varphi^a$ by the spinless fermions $\psi_p$ and the complex bosonic
spinors $z_\alpha$.  A crucial feature of the resulting effective Lagrangian for the $\psi_{p}$ and $z_{\alpha}$ is that it is
invariant under a local U(1) gauge transformation. This follows from the invariance of Eqs.~(\ref{cab}) and (\ref{nz})
under
\begin{eqnarray}
z_{i\alpha} &\rightarrow &z_{i\alpha} e^{i \phi_i} \nonumber \\
\psi_{i p} & \rightarrow & \psi_{i p} e^{- i p \phi_i}
\end{eqnarray}
where $\phi_i$ has an arbitrary dependence on space and time. Note that the $\psi_{i p}$ have opposite charges
$p=\pm 1$. Associated with this U(1) gauge invariance, we will introduce an internal dynamical gauge field $A_\mu = (A_\tau, {\bm A})$
in constructing the effective theory.

We can now insert Eqs.~(\ref{cab}) and (\ref{nz}) into Eqs.~(\ref{hsdw}) and (\ref{ln}) and obtain the desired effective
theory of fluctuating Fermi pockets. As noted earlier, we will assume that the $z_{i \alpha}$ are slowly varying, but allow the 
fermion fields $\psi_{i p}$ to have an arbitrary dependence on spacetime.
First, from Eq.~(\ref{ln}), we obtain, by a familiar method \cite{review}, the CP$^1$ model for the $z_\alpha$:
\begin{equation}
\mathcal{L}_z = \frac{1}{g} \left[ | (\partial_\tau - i A_\tau) z_\alpha |^2 + v^2 |({\bm \nabla} - i {\bm A} ) z_\alpha|^2 \right] \label{lz}
\end{equation}
The fermion Lagrangian $\mathcal{L}_F$ in Eq.~(\ref{hsdw}) yields some interesting structure. The hopping term can be written as
\begin{eqnarray}
&& - \sum_{i<j} t_{ij} \Biggl[   \bigl( z_{i \alpha}^\ast z_{j \alpha} \bigr) \left( \psi^\dagger_{i+} \psi_{j+} + \psi^\dagger_{j-} \psi_{i-} \right)
\nonumber \\
&&~~~~~~~~~+ \bigl( z_{j \alpha}^\ast z_{i \alpha} \bigr) \left( \psi^\dagger_{i-} \psi_{j-} + \psi^\dagger_{j+} \psi_{i+} \right) \nonumber \\
&&~~~~~~~~~+\bigl( \varepsilon^{\alpha\beta} z_{j \alpha}^\ast z_{i \beta}^\ast \bigr) \left( \psi^\dagger_{i+} \psi_{j-}
- \psi_{j+}^\dagger \psi_{i-} \right) \nonumber \\
&&~~~~~~~~~+ \bigl( \varepsilon^{\alpha\beta} z_{i \alpha} z_{j \beta} \bigr) \left( \psi^\dagger_{i-} \psi_{j+} 
- \psi^\dagger_{j-} \psi_{i+} \right)
\Biggr] \label{hop}
\end{eqnarray}
Now, from the derivation of the CP$^1$ model \cite{review} we know that
\begin{equation}
z_{i \alpha}^\ast z_{j \alpha} \approx e^{i A_{ij}}
\end{equation}
and this is easily incorporated into the first two terms in Eq.~(\ref{hop}), yielding terms which are gauge invariant.
Then for the fermion sector, we have the Lagrangian
\begin{eqnarray}
\mathcal{L}_\psi &=& \sum_{p=\pm1} \sum_{i} \psi_{ip}^\dagger
\Bigl( \partial_\tau - \mu + i p A_{\tau}  - p f_i \Bigr) \psi_{ip}  \nonumber \\
&~&- \sum_{p=\pm 1} \sum_{i<j} t_{ij} \left( e^{i p A_{ij}} \psi_{ip}^\dagger \psi_{jp} + e^{-i p A_{ij}} \psi_{jp}^\dagger \psi_{ip} \right)
\label{lpsi}
\end{eqnarray}
For $A_\mu = 0$, $\mathcal{L}_\psi$ describes the band structure in terms of the Fermi pockets; and 
the interactions arise from the minimal coupling to the $A_\mu$ gauge field.
Finally, we need to consider the last two terms in Eq.~(\ref{hop}). These are the analog of the `Shraiman-Siggia' couplings \cite{shraiman}; this
evident from their form expanded to leading order
in the derivative of the $z_\alpha$:
\begin{equation}
\mathcal{L}_{ss} =  \int_{{\bm k},{\bm p},{\bm q}} \left[ {\bm p} \cdot \frac{\partial \varepsilon ({\bm k})}{\partial {\bm k}} \right] z_{\downarrow} ({\bm q}-{\bm p}/2) z_{\uparrow} ({\bm q}+{\bm p}/2) \psi^\dagger_- ({\bm k} + {\bm q}) \psi_+ ({\bm k} - {\bm q})  + \mbox{c.c.} \label{lss}
\end{equation}
where $\varepsilon ({\bm k})$ is the single particle dispersion of the large Fermi surface state:
\begin{equation}
\varepsilon ({\bm k} ) = - \sum_j t_{ij} e^{i {\bm k} \cdot ({\bm r}_j - {\bm r}_i)}.
\end{equation}

The Lagrangian $\mathcal{L}_z+\mathcal{L}_\psi + \mathcal{L}_{ss}$ in Eqs.~(\ref{lz}), (\ref{lpsi}), and (\ref{lss}) 
is our final and general form of the U(1) gauge theory
of the fluctuating spin density wave state. Note that it is applicable for an arbitrary band structure $\varepsilon ({\bm k})$ and for
an arbitrary SDW wavevector ${\bm K}$: thus we have satisfied points 2 and 3 in Section~\ref{sec:intro}. After diagonalizing the band structure
of $\mathcal{L}_\psi$ in Eq.~(\ref{lpsi}), and projecting to the resulting lowest energy electron and hole pocket states, the present
model reduces to those considered in our previous work \cite{rkk1,rkk2,rkk3,gs,moon}. Note also that this model is specialized
to the fluctuating pocket state, and there is no natural way of restoring the large Fermi surface: the coupling to the local SDW order
in Eq.~(\ref{lpsi}) has a fixed magnitude set by the $f_i$.

The phase diagram of the theory  $\mathcal{L}_z+\mathcal{L}_\psi$ contains phases (A) and (B) in Fig.~\ref{phase} appearing in 
Section~\ref{sec:su2}. These are the Fermi liquid SDW (with $\langle z_\alpha \rangle \neq 0$) and `algebraic charge liquid' (with
$\langle z_\alpha \rangle = 0$) phases respectively, and will be discussed further in Section~\ref{sec:su2}.

\section{SU(2) gauge theory}
\label{sec:su2}

The structure of the Shraiman-Siggia term, $\mathcal{L}_{ss}$, exposes a shortcoming of the U(1) theory. In the gradient
expansion, this term is of the same order as the U(1) gauge field term in $\mathcal{L}_\psi$ in Eq.~(\ref{lpsi}). It is only the collinear
nature of the local spin order which imposes the U(1) gauge structure, while $\mathcal{L}_{ss}$ 
is associated with spiral spin correlations \cite{shraiman}. However, once we are in the large Fermi surface state, the memory
of the collinear spin correlations should disappear. Thus,
if we are to recover the large Fermi surface state, we will have to treat all the terms in Eq.~(\ref{hop}) at
an equal footing.

To fix this problem, we 
note that the parameterization in Eq.~(\ref{cab}) actually introduces a SU(2) gauge invariance under which
\begin{equation}
R \rightarrow R U^\dagger~~~;~~~\psi \rightarrow U \psi.
\end{equation}
Thus the SU(2) gauge transformation acts on the second index of $R$ (denoted by $p$), while the ordinary SU(2) 
spin rotation symmetry acts on the left index of $R$ (denoted by $\alpha$). We will distinguish the SU(2) gauge
and SU(2) spin rotation invariances by using the symbols $p,p'$ and $\alpha,\beta$ for their respective spinor
indices.
Using the parameterization in Eq.~(\ref{cab}) on the coupling between the SDW order and the fermions 
in Eq.~(\ref{hsdw}), we find that it can be written as
\begin{equation}
\sum_i N^\ell_i f_i \psi^\dagger_{i p} \sigma^\ell_{pp'} \psi_{i p'} \label{yukawa}
\end{equation}
where we have introduced a field $N^\ell_i$, which transforms as a adjoint under the SU(2) gauge transformation.
Again, we will distinguish the SU(2) gauge
and SU(2) spin rotation invariances by using the symbols $\ell=x,y,z$ and $a$ for their respective adjoint
indices. From Eq.~(\ref{hsdw}) we find that
\begin{equation}
N^\ell_i = \frac{1}{2} \varphi_i^a \, \mbox{Tr} \left( \sigma^a R^i \sigma^\ell R^{i\dagger} \right). \label{rz2}
\end{equation}
This relationship is equivalent to 
\begin{equation}
\varphi^a_i = \frac{1}{2} N_i^\ell \, \mbox{Tr} \left( \sigma^a R^i \sigma^\ell R^{i\dagger} \right). \label{rz3}
\end{equation}
Only for $N^\ell = (0,0,1)$ does
Eq.~(\ref{rz3}) yield the relation Eq.~(\ref{nz}).

Let us now summarize the structure of our effective gauge theory.
The theory has  SU(2)$_{\rm gauge} \otimes$SU(2)$_{\rm spin} \otimes$U(1)$_{\rm em~charge}$
invariance, along with additional constraints from the lattice space group symmetry. Its matter content is:
\begin{itemize}
\item  A fermion $\psi$ transforming as $({\bm 2}, {\bm 1}, 1)$, and with dispersion $\varepsilon ({\bm k})$
from the underlying lattice band structure.
\item A relativistic SU(2) matrix field $R$ (with $R^\dagger R = 1$) 
transforming as $(\bar{\bm 2}, {\bm 2}, 0)$, representing the local orientational fluctuations of the 
SDW order. The notation indicates that $R$ transforms under SU(2)$_{\rm gauge}$ under right multiplication,
and under SU(2)$_{\rm spin}$ under left multiplication.
\item  A relativistic real scalar $N$ transforming as $({\bm 3}, {\bm 1}, 0)$, measuring the local
SDW amplitude.
\end{itemize}
The symmetries allow a Yukawa coupling between $N$ and $\psi$, which is just the coupling in 
Eq.~(\ref{yukawa}). Note that this coupling has a space dependence $\sim e^{i {\bm K} \cdot {\bm r}}$,
which can understood to be a consequence of the non-trivial transformation of the SDW order parameter, and hence
of $N^\ell$, under the square lattice space group.

Now, we can introduce a SU(2) gauge field $A^\ell_\mu = (A^\ell_\tau, {\bm A}^\ell)$, and use the above constraints to write down
our low energy effective action for the SDW fluctuations. The fields $R$ and $N^\ell$ will have conventional kinetic energy
terms familiar from relativistic non-Abelian gauge theory, similar to those in $\mathcal{L}_z$:
\begin{eqnarray}
\mathcal{L}_R &=& 
 \frac{1}{g} \left[ | \partial_\tau R_{\alpha p} - i A^\ell_\tau R_{\alpha p'} \sigma^\ell_{p'p}  |^2 + v^2 |  {\bm \nabla} R_{\alpha p} - i {\bm A}^\ell R_{\alpha p'} \sigma^\ell_{p'p} |^2 \right]; \label{lrn} \\
\mathcal{L}_N &=& \left( \partial_\tau N^\ell - 2 i \epsilon_{\ell m n} A^m_\tau N^n \right)^2 + \widetilde{v}^2 \left( {\bm \nabla} N^\ell - 2 i \epsilon_{\ell m n} {\bm A}^m N^n \right)^2 + s ( N^\ell )^2 + u ( ( N^\ell )^2 )^2, \nonumber
\end{eqnarray}
where $g$, $r$ and $u$ are couplings which tune the strength of the SDW fluctuations.
For the fermions, $\psi_{ip}$, we now have a lattice Lagrangian which is similar to Eq.~(\ref{lpsi}), but invariant SU(2) gauge transformations
\begin{eqnarray}
\widetilde{\mathcal{L}}_\psi &=&  \sum_{i} \psi_{ip}^\dagger
\Bigl[ (\partial_\tau - \mu) \delta_{pp'} + i A^\ell_{\tau} \sigma^\ell_{pp'}  - f_i N^\ell_i \sigma^\ell_{pp'} \Bigr] \psi_{ip'}  \nonumber \\
&~&-  \sum_{i,j} t_{ij}  \psi_{ip}^\dagger \left( e^{i \sigma^\ell {\bf A}^\ell \cdot ({\bf r}_i - {\bf r}_j) } \right)_{pp'} \psi_{jp'}. 
\label{lpsi2}
\end{eqnarray}
Apart from the generalization of the U(1) gauge field to SU(2), the main difference from Eq.~(\ref{lpsi}) is that the coupling
of the fermions to the SDW order has a magnitude determined by the field $N^\ell$. Thus a phase in which $N^\ell$ fluctuates
near zero can have a large Fermi surface given by the underlying band structure.

\begin{figure}
\includegraphics*[width=5in]{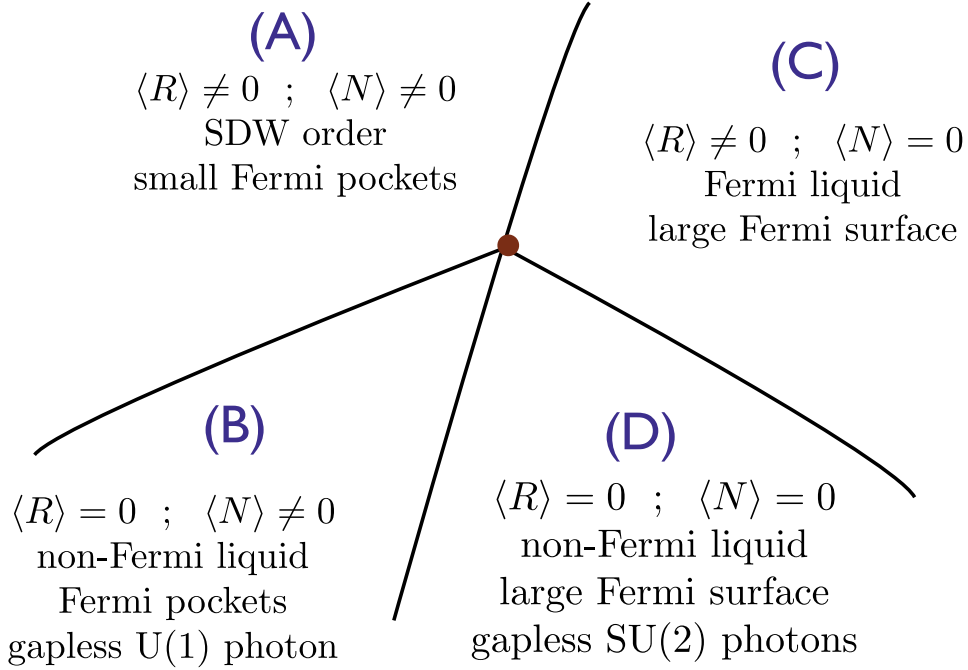}
\caption{Mean field  phase diagram of SU(2) gauge theory. The phases (A) and (B) are also obtained within the U(1) gauge theory of
Section~\ref{sec:u1}, as is the transition between them. The phases (C) and (D), 
and all other transitions, require the SU(2) gauge theory. The Fermi liquid phases (A) and (C) also appear as the non-superconducting 
ground states in Fig.~\ref{figglobal}
at large $H$. The non-Fermi liquid phases (B) and (D) could appear as intermediate phases at $T=0$ and large $H$
(they are not shown in Fig.~\ref{figglobal}). We have
argued previously \cite{rkk2,gs,moon} that phase (B) describes the crossovers at $T>0$, $H=0$ for $x<x_m$ in Fig.~\ref{figglobal}.
We suggest here that phase (D) may be useful in the description of the strange metal in Fig.~\ref{figglobal}; alternatively, as is indicated
in Fig.~\ref{figglobal}, the strange metal may simply be a reflection of the quantum criticality of the transition between the Fermi liquid 
phases (A) and (C).
 }
\label{phase}
\end{figure}

We are now in a position to discuss the mean-field phase diagram of the SU(2) gauge theory $\mathcal{L}_R + \mathcal{L}_N + \widetilde{\mathcal{L}}_\psi$. Initially, we take a simple-minded approach by allowing Higgs condensates of one or both of the bosonic fields $R$ and $N$. This allows 4 possible phases which are sketched in Fig.~\ref{phase}.
As will become clear below, there are no other phases that can generically be expected.

Note that a phase which breaks SU(2) spin rotation invariance requires condensation of both $R$ and $N$: this is clear from Eq.~(\ref{rz3})
which shows that both condensates are required for a non-zero SDW order parameter $\varphi$. The other three phases
preserve SU(2) spin symmetry, and we now discuss the various possibilities.
\begin{enumerate}[(A)]
\item
The Higgs phase, noted above, with $\langle R \rangle \neq 0$ and $\langle N \rangle \neq 0$. 
In this case both spin rotation symmetry and SU(2) gauge symmetry is broken, and there are no gapless gauge bosons. 
So this phase is a Fermi liquid, and is the conventional SDW state with Fermi pockets. It appears in Fig.~\ref{figglobal} as the ground
state at large $H$ and $x<x_m$.
\item
Higgs phase with $\langle N \rangle \neq 0$, but spin SU(2) symmetry preserved because $\langle R \rangle = 0$.
We can always choose $N^\ell \sim (0,0,1)$ by a gauge transformation, and we then find that a U(1) subgroup of the SU(2)
gauge group remains unbroken because the $A^z_\mu$ photon remains gapless. Thus at low energies we
have  a U(1) gauge theory, and
the fermion Lagrangian
$\widetilde{\mathcal{L}}_\psi$ in Eq.~(\ref{lpsi2}) becomes equivalent to $\mathcal{L}_\psi$ in Eq.~(\ref{lpsi}). Thus this phase
reduces to the non-Fermi liquid phase of the U(1) gauge theory, which is the `algebraic charge liquid' of Refs.~\onlinecite{rkk2,rkk3,gs,moon}.
The fermions have a Fermi pocket dispersion, and the gapless U(1) photon produces non-Fermi liquid behavior
at the Fermi surface. This phase is not shown in Fig.~\ref{figglobal}, but it is a possible ground state for large $H$ near $x=x_m$.
This phase has also played a key role in our previous studies \cite{rkk2,gs,moon} of the physics
at $H=0$, $x< x_m$, and $T \gtrsim T_c$. 
\item
SU(2) confining phase: this is the Fermi liquid with the large Fermi surface. We can also think of this phase
as the Higgs phase of a condensate which transforms as a SU(2) fundamental {\em i.e.\/} $\langle R\rangle \neq 0$.
Also this phase should have $\langle N \rangle =0$ to preserve spin rotation invariance. Note that the condensate of $R$
alone does not break SU(2) spin invariance, because the condensate can be rotated into an arbitrary direction by a SU(2)
gauge transformation. This phase appears in Fig.~\ref{figglobal} as the ground
state at large $H$ and $x>x_m$.
\item 
A novel phase with no fields condensed. This is also an algebraic charge liquid, but there are a triplet of gapless SU(2) photons.
The fermions have a large Fermi surface dispersion, with non-Fermi liquid behavior along the Fermi surface; this is in contrast
to the small Fermi pockets in phase (B). This phase is not shown in Fig.~\ref{figglobal}, but like phase (B), 
it is a possible ground state for large $H$ near $x=x_m$. The existence of critical behavior across the entire large Fermi
surface, with no pocket-like structures, 
also makes this state a possible starting point for describing the strange metal phase of Fig.~\ref{figglobal}.
\end{enumerate}

Going beyond these mean-field considerations, it is clear that all these phases are unstable to pairing and 
a superconducting instability \cite{rkk2,rkk3,gs,moon}.
However, it is still meaningful to ask whether the metallic states and critical points remain stable, after superconductivity
has been suppressed {\em e.g.\/} by an applied magnetic field.

Note that the discussion for phases (A) and (B) reduces to that
in the U(1) theory. The stability of these phases was established in Ref.~\onlinecite{rkk3}, and it was noted that
the (A)-(B) transition was in the O(4) universality class.

It is clear that the Fermi liquid state (C) is stable. Let us then consider the transition from state (C) to the SDW state (A).
Note that neither of these states have gapless gauge bosons, and both are conventional Fermi liquids. Indeed, the order parameter
for the (A)-(C) transition is the vector $N^\ell$; we can always choose the gauge $R=1$, and then this order parameter is seen 
from Eq.~(\ref{rz3}) to be the 
conventional SDW order parameter $\varphi^a$. It should now be clear that the effective theory for the (A)-(C) transition reduces
to the well-known Hertz theory \cite{vrmp}. It is quite remarkable that after the detour to fractionalized degrees of freedom,
our theory has produced the same answer as that expected from ``Landau-Ginzburg'' reasoning. We should note that key open question
remain for the Hertz SDW transition in two spatial dimensions: Abanov and Chubukov \cite{abchub} have shown that the
theory has an infinite number of marginal operators, and the nature of the quantum critical point remains open.

Finally, we turn to the issue of the stability of the non-Fermi
liquid phase (D) with gapless SU(2) photons. Corresponding issues have been discussed
in the literature \cite{schafer} for the non-Abelian gauge theory of quark matter in three
spatial dimensions, and we discuss the two dimensional case here.
We recall that in the
presence of the Fermi surface, the longitudinal component of the
SU(2) gauge-boson is Debye screened, leaving only the transverse
component at low energies. This transverse component is, in turn,
Landau damped, so that the gauge sector of the Lagrangian has a
dynamical critical exponent $z = 3$, rendering the bare
self-interactions of the gauge bosons irrelevant. Moreover, a
gauge-boson can only interact efficiently with the patch of the
Fermi surface that is tangent to its momentum. This interaction is
singular enough to destroy the Fermi liquid: at one loop the
fermion acquires a self-energy that scales as $\omega^{2/3}$. The
form of the one loop effective action leads one to hypothesize an
anisotropic z = 3 scaling\cite{Polchinski} under which $\omega
\sim k_{\parallel}^3$, $k_{\perp} \sim k^2_{\parallel}$, where
$k_{\parallel}$ and $k_{\perp}$ are the components of the fermion
momentum parallel and perpendicular to the Fermi surface. The
self-interactions of the gauge bosons are irrelevant under this
scaling as well.

We would like to caution the reader that the above discussion is
based on a one-loop analysis. Higher loop diagrams can still cause
nonperturbative effects at low energy. In Ref.~\onlinecite{sslee}, in
order to introduce a small parameter for expansion, the author
studied a simplified situation with $N$ copies of identical Fermi
patches coupled with a U(1) gauge boson. In the large-$N$ limit
the gauge coupling is not flowing under RG and the system has a
deconfined phase, which obeys the same scaling as the one loop
result. This conclusion carries over directly to the SU(2) case.
However, when there is a full Fermi surface even the large-$N$
limit becomes much more complicated. We will leave a more detailed
investigation of the stability of the phase (D) to future study.

\section{Conclusions}

The physics of doped antiferromagnets has been a subject of intense study since the discovery of the 
cuprate superconductors. At low doping in the ordered antiferromagnet, we obtain metallic Fermi liquid states
with Fermi pockets. Much theoretical work has focused on the fate of these pockets when the quantum
and thermal fluctuations of the antiferromagnetic (or SDW) order start to increase. These issues
have usually been addressed \cite{shraiman,schulz,wenholon,leeholon,shankar,ioffew,rkk1,rkk2,rkk3,gs} 
using the strong-coupling perspective of the $t$-$J$ model, appropriate
to a doped Mott insulator. The claimed discovery of electron pockets in the hole-doped cuprates \cite{louis},
suggested \cite{moon,qcnp} that a weak-coupling perspective may also be useful. Here we used the ``spin-fermion''
model \cite{spinfermion} to provide an efficient derivation of the same effective U(1) gauge theory that
appears in the strong-coupling approach. Our new approach had the added advantages of being applicable
to arbitrary band structures and ordering wavevectors. The main idea  \cite{shraiman,schulz,schrieffer,gs,moon}  behind our 
analysis was to transform
the electron spin polarization to a rotating frame of reference determined by the local orientation of the SDW order.

In the second part of the paper we addressed the transition from the Fermi pocket SDW state to the large doping
Fermi liquid with a large Fermi surface. We showed that such a transition required embedding the U(1) gauge
theory into a SU(2) gauge theory. Unexpectedly, the SU(2) gauge theory displayed a direct transition
between Fermi liquid states with and without SDW order, which was described by the same effective
low energy theory as that obtained by Hertz \cite{vrmp} and Abanov and Chubukov \cite{abchub}. 
Thus our formalism, expressed in terms of fractionalized
degrees of freedom, can also efficiently describe the transition between confining states. The SU(2) theory
also allowed for intermediate fractionalized phases between the Fermi liquid states with and without SDW order,
as is shown in Fig.~\ref{phase}.

In the cuprates, the possibility remains open that the fractionalized phases of Fig.~\ref{phase} are present
as stable $T=0$ phases in high magnetic fields between the under- and over-doped regimes in Fig.~\ref{figglobal}
(they are not shown in Fig.~\ref{figglobal}).
Irrespective of whether they are present at $T=0$, the fractionalized phases provide 
a useful description of the finite temperature crossovers. We have previously described \cite{rkk2,gs,moon,qcnp}
the use of the U(1) fractionalized phase (B) in the underdoped regime: we showed that it reproduces all qualitative
features of the phase diagram in Fig.~\ref{figglobal} for $x<x_m$, including the crucial shift between $x_s$ and $x_m$. 
With the improved 
formalism presented here for arbitrary band structure and ordering wavevector, we hope that more
quantitative tests of this theory will be possible, especially for the fermion spectral functions in the underdoped regime \cite{rkk2}. 
Finally, the novel SU(2) fractionalized phase (D) offers a possible
framework for developing a theory of the strange metal; such a description would be an alternative to the possibility \cite{qcnp}
indicated in Fig~\ref{figglobal}: the strange metal reflects the quantum criticality between the Fermi liquid phases (A) and (C).

\acknowledgements

We thank A.~Chubukov, H.~Liu, J.~McGreevy, K.~Rajagopal, A.~Rosch, D.~Scalapino, T.~Senthil, and Sung-Sik Lee for valuable discussions.
This research was supported by the National Science Foundation under grant DMR-0757145, by the FQXi
foundation, and by a MURI grant from AFOSR. The research at KITP was supported in 
part by the National Science Foundation under Grant No. PHY05-51164.

\end{document}